\documentclass[%
 reprint,
 amsmath,amssymb,
 aps,
]{revtex4-1}

\usepackage{graphicx}% Include figure files
\usepackage{dcolumn}% Align table columns on decimal point
\usepackage{bm}% bold math
\begin{document}

%\preprint{APS/123-QED}

\title{Neutrino quantum decoherence engendered by neutrino radiative decay}% Force line breaks with \\

\author{Konstantin Stankevich}
\email{kl.stankevich@physics.msu.ru}
\affiliation{%
	Department of Theoretical Physics, Faculty of Physics,
Lomonosov Moscow State University, 119992 Moscow, Russia
}


\author{Alexander Studenikin}
\email{studenik@srd.sinp.msu.ru}
\affiliation{
	Department of Theoretical Physics, Faculty of Physics,
	Lomonosov Moscow State University, Moscow 119991, Russia
}
\affiliation{Dzhelepov Laboratory of Nuclear Problems,
	Joint Institute for Nuclear Research,
	Dubna 141980, Moscow Region, Russia
}%

\date{\today}% It is always \today, today,

\begin{abstract}
	A new theoretical framework, based on the quantum field theory of open systems applied to neutrinos, has been developed to describe the neutrino evolution in external environments accounting for the effect of the neutrino quantum decoherence. The developed new approach enables one to obtain
the explicit expressions of the decoherence and relaxation parameters that 
account for a particular process, in which the neutrino participates, and also
for the characteristics
of an external environment and of the neutrino itself, including the neutrino energy.  We have used this approach to consider a new mechanism of the neutrino quantum decoherence engendered by the neutrino radiative decay to photons and dark photons in an astrophysical environment. The importance of the performed studies  is highlighted by the prospects of the forthcoming new large volume neutrino detectors that will provide new frontier in high-statistics measurements of neutrino fluxes from supernovae.

\end{abstract}

\pacs{Valid PACS appear here}% PACS, the Physics and Astronomy
\maketitle

%\tableofcontents

%%%%%%%%%%%%%%%%%%%%%%%%%%%%%%%%%%%%%%%%%%%%%%%%%%%%%%%%%%%%%%%%%%
\section{Introduction}
%%%%%%%%%%%%%%%%%%%%%%%%%%%%%%%%%%%%%%%%%%%%%%%%%%%%%%%%%%%%%%%%%%
Half a century ago Gribov and Pontecorvo derived \cite{Gribov:1968kq} the first analytical expression for the neutrino oscillation probability that has opened a new era in the theoretical and experimental studies of the neutrino oscillation phenomenon. The neutrino oscillation patterns can be modified by neutrino interactions with external environments including electromagnetic fields that can influence on neutrinos in the case neutrinos have nonzero electromagnetic properties \cite{Giunti_Studenik}. The phenomenon of neutrino oscillations can proceed only in the case of the coherent superposition of neutrino mass states. An external environment can modify a neutrino evolution in a way that conditions for the coherent superposition of neutrino mass states are violated. Such a violation is called quantum decoherence of neutrino states and leads to the suppression of flavor neutrino oscillations. It should be noted that the quantum neutrino decoherence differs from  the standard neutrino decoherence that appears due to separation of neutrino wave packets, the effect that is not considered below.

The quantum neutrino decoherence has attracted a growing interest during the last 15 years. Within reasonable amount of the performed studies the method based on the Lindblad master equation \cite{Lindblad, Gorini_Kossakowski} for describing neutrino evolution has been used. This approach is usually considered as the most general one that gives a possibility to study neutrino quantum  decoherence as a consequence of standard and nonstandard interactions of a neutrino system with an external environments \cite{Farzan_Schwetz_Smirnov,Lisi_Marrone,Barenboim_Mavromato,Barenboim_Mavromatos2,	 Benatti_Floreanini,Oliveira2014,Oliveira2016,
Balieiro_Guzzo,Joao_Coelho,Joao_Coelho2,Capolupo_Giampalo}.

The Lindblad master equation can be written in the following form (see, for instance, \cite{Joao_Coelho})

\begin{equation}
\dfrac{ \partial \rho_\nu (t)}{\partial t} = - i \left[ H_S, \rho_\nu (t) \right] + D\left[ \rho_\nu \right]
,
\label{Lindblad}
\end{equation}
where $\rho_\nu$ is the density matrix that describes the neutrino evolution, $H_S$ is the Hamiltonian and the dissipation term (or dissipator) is given by

\begin{equation}
D \left[ \rho_\nu(t) \right] = \dfrac 1 2 \sum^{N^2-1}_{k=1} \left[ V_k, \rho_\nu V^\dag_k \right] + \left[ V_k \rho_\nu, V_k^\dag \right]
,
\end{equation}
where $V_k$ are dissipative operators that arises from interaction between the neutrino system and the external environment, these operators act only on the N-dimensional $\rho_\nu$ space.

For two-neutrino approximation operators $V_k$ can be expanded by Pauli matrices $O = a_\mu \sigma_\mu$, where $\sigma_\mu$ are composed by an identity matrix and the Pauli matrices. In this case the equation (\ref{Lindblad}) can be written as

\begin{equation}
\dfrac{\partial \rho_k (t)}{\partial t} \sigma_k = 2 \epsilon_{ijk} H_i \rho_j (t) \sigma_k + D_{kl} \rho_l(t) \sigma_k
\label{LindbladEq}
,
\end{equation}
where
\begin{equation}
D_{ll} = - diag \{ \Gamma_1, \Gamma_1, \Gamma_2 \}
\label{par}
,
\end{equation}
 and $\Gamma_1$ and $\Gamma_2$ are the parameters that describe two dissipative effects: 1) the decoherence effect, and 2) the relaxation effect, respectively. In the case of the energy conservation in the neutrino system  there is an additional requirement on a dissipative operators \cite{Oliveira2016},
\begin{equation}
[H_S, V_k]=0
.
\end{equation}
In this case the relaxation parameter is equal to zero $\Gamma_2=0$.

In the approach that uses the Lindblad master equation the quantum decoherence is described by the free dissipative parameters (decoherence parameter $\Gamma_1$ and relaxation parameter $\Gamma_2$) that can be constrained (or determined) by the  experimental data on neutrino fluxes. Currently the  long-baseline neutrino experiments provide constraints on the order of $\Gamma_1 \sim 10^{-24}$ GeV \cite{Coloma} on the decoherence parameter .

There are peculiar limitations in implementation of the Lindblad approach to description of the neutrino quantum decoherence.
This is because within this approach it is not possible to calculate the energy dependence of the decoherence and relaxation parameters. On the contrary, the energy dependence is fixed {\it ad hoc}, at the same time the obtained constraints on the parameters are neutrino energy dependent. An interesting and detailed recent discussion of this issue and the corresponding references to the existed related studies can be found in \cite{Joao_Coelho2}.

It should be noted that at present there is no general theoretical approach to description of the neutrino quantum decoherence phenomenon that might be used in the  derivation of the explicit expressions for the decoherence and  relaxation parameters (\ref{par}) accounting for a concrete type of the neutrino interaction with the environment. Probably, the only exception are provided in  \cite{Burgess_Michaud, Benatti_Florianini_2} where the quantum decoherence is calculated as a consequence of the neutrino interaction with the matter fluctuations. However, the developed approach is not general and it can not be used in the case of other neutrino decoherence mechanisms.

In this paper, we propose and develop a new theoretical framework, based on the quantum field theory of open systems \cite{Breuer_Pettrucione}, for the neutrino evolution in an external environment. Here below we implement the proposed approach to the consideration of a new mechanism of the neutrino quantum decoherence that appears due to neutrino radiative decay in the thermal background of electrons and photons. Within our approach we have obtained the explicit expressions of the decoherence and relaxation parameters as functions of the characteristics of an external environment and also of the neutrino energy. We also apply the developed approach to the study a possible influence of dark photons on the neutrino quantum decoherence.

Note that the influence of the neutrino radiative decay, as well as the
neutrino interaction with dark photons,  on the neutrino oscillation phenomenon
and the corresponding contributions to the neutrino quantum decoherence are
considered for the first time. The developed theoretical framework provides a  general basis for the detailed description of
the neutrino quantum decoherence due to different neutrino interactions
with external environments.

%%%%%%%%%%%%%%%%%%%%%%%%%%%%%%%%%%%%%%%%%%%%%%%%%%%%%%%%%%%%%
\section{Neutrino quantum decoherence engendered by neutrino radiative decay}
%%%%%%%%%%%%%%%%%%%%%%%%%%%%%%%%%%%%%%%%%%%%%%%%%%%%%%%%%%%%%

For description of the neutrino decoherence we use the formalism of quantum electrodynamics of open systems which was used in \cite{Breuer_Pettrucione} for evolution of electrons. We start with the quantum Liouville equation for the density matrix of a system composed of neutrinos and an electromagnetic field

\begin{equation}
\frac{\partial}{\partial t} \rho =  - i \int d^3 x\left[ H(x),\rho \right]
\label{rho}
,
\end{equation}
where

\begin{equation}
H(x) = H_\nu(x) + H_{int}(x)
\end{equation}
is the Hamiltonian density of the system in the interaction picture. $H_\nu(x)$ is the Hamiltonian density of the neutrino system and $H_{int}(x)$ describes interaction between neutrino and the electromagnetic field
\begin{equation}
H_{int} (x) = j_\alpha (x) A^\alpha (x)
,
\end{equation}
where $j_\alpha (x)$ is the current density of neutrino and $A_\alpha$ is the electromagnetic field.

\begin{figure}[t!]
	\centering
	\includegraphics[width=1\linewidth]{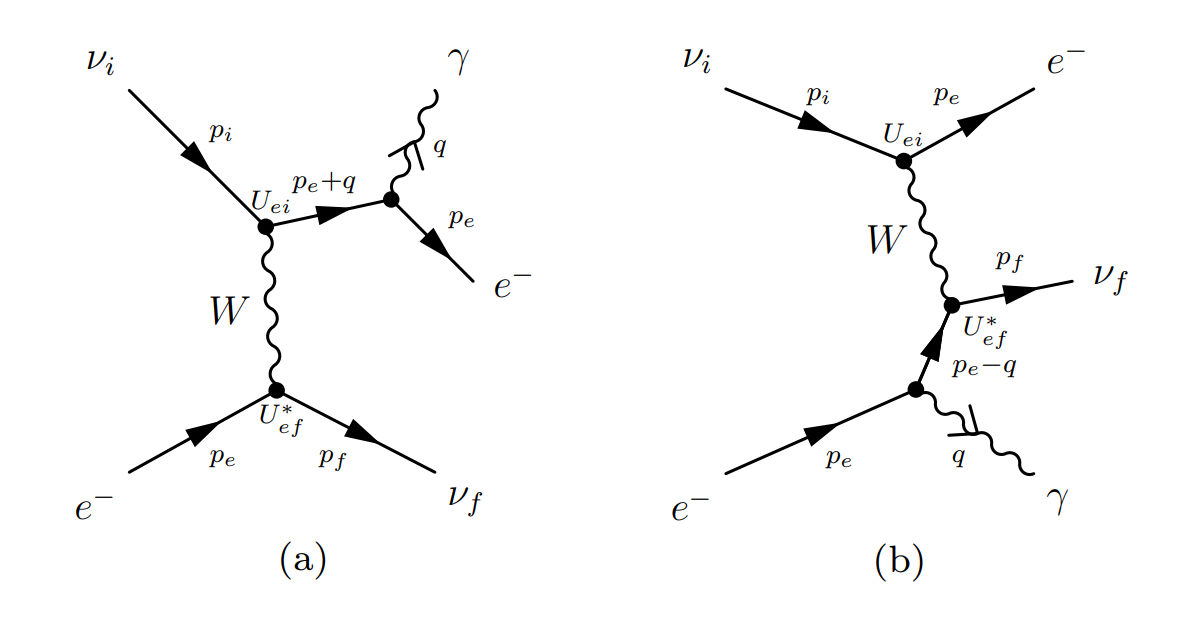}
	\caption{\label{Feinman} Feynman diagrams of neutrino radiative decay in matter (see, for instance, \cite{Giunti_Studenik}).}
\end{figure}

Here below within our new approach we propose and study a new mechanism of the quantum decoherence that appears due to the neutrino radiative decay (see Fig.\ref{Feinman}). In this case the current density can be expressed in the following form \cite{Olivio_Nieves_Pal}

\begin{equation}
j_\alpha (x) = \overline{\nu}_i(x)\Gamma_\alpha \nu_j(x)
\label{current0}
,
\end{equation}
where $\nu_i(x)$ is the neutrino field with mass $m_i$ ($i,j=1,2,3,4$). Here below, we also include $eV$ sterile neutrino state $\nu_4$ into the consideration. The oscillation parameters for a sterile neutrino we take from \cite{Knee:2018rvj}.

In \cite{Olivio_Nieves_Pal} the radiative decay is considered between vacuum neutrino mass states that are not stationary in the electron media. In our studies we are working within the formalism that we have used for the quantum theory of the spin light of neutrino \cite{Studenikin:2004gg, Grigorev:2005sw, Grigoriev:2017wff} where transitions occur between the neutrino stationary states in matter.

In (\ref{current0}) $\Gamma_\alpha$ is an effective electromagnetic vertex

\begin{equation}
\Gamma_\alpha = U^*_{ei} U_{ej} \tau_{\alpha\beta} \gamma^\beta L
,
\label{vertex}
\end{equation}
where $U$ is the lepton mixing matrix and $L = \frac 1 2 (1-\gamma_5)$ is the projection operator for the left-handed fermions. It is supposed that the four-velocity of the center of mass of the electron background is at rest. In the case of a nonrelativistic (NR) background $\tau_{\alpha\beta}$ can be expressed as \cite{Olivio_Nieves_Pal}

\begin{equation}
\tau_{\alpha\beta}^{NR} =\tau^{NR} P_{\alpha\beta}  = - \sqrt{2} \frac{e G_F n_e} {m_e} P_{\alpha\beta}
.
\end{equation}
The tensor

\begin{equation}
P_{\alpha\beta} = \delta_{\alpha\beta} - \frac{k_\alpha k_\beta}{|\vec k|^2}
\end{equation}
 is a projector onto the transverse component in $k$-space, $e$, $m_e$ and $n_e$ are the charge and mass of an electron and the medium number density, respectively.

In the extreme relativistic (ER) case, when the temperature of the background electrons $T\gg m_e$, one gets

\begin{equation}
\tau_{\alpha\beta}^{ER} =\tau^{ER} P_{\alpha\beta} = - \frac{e G_F T^2}{2 \sqrt{2}} P_{\alpha\beta}
\label{ER}
.
\end{equation}
The corresponding expression for the degenerate electron gas of neutron stars reads

\begin{equation}
\tau_{\alpha\beta}^{Deg} =\tau^{Deg} P_{\alpha\beta} = - \dfrac{\sqrt 2 e G_F}{4} \left(\dfrac{3 n_e}{\pi}\right)^{2/3} P_{\alpha\beta}
.
\end{equation}

Using the exact free neutrino mass states spinors

\begin{equation}
\nu_i = C_i \sqrt{\dfrac{E_i+m_i}{2 E_i}}
\left(
\begin{matrix}
u_i\\
\frac{\boldsymbol{\sigma} \textbf{p}_i} {E_i+m_i}u_i
\end{matrix}
\right)
e^{i \textbf{p}_i\textbf{x}}
,
\end{equation}
where $u_i$ are the two component spinors, we express the current density (\ref{current0}) in the form

\begin{equation}
j_3 = 2 U^*_{ei} U^*_{ej} \tau
\left(
\begin{matrix}
0 & 1 \\
1 & 0
\end{matrix}
\right)
,
\label{current}
\end{equation}
where for different cases $\tau$ stands for  $\tau^{ER}$, $\tau^{NR}$ or $\tau^{Deg}$. Here we use the analogous calculations to those performed in \cite{Fabbricatore_Grigoriev,Pustoshn}.

Note that only the third component of the current is responsible for the neutrino decay. It is convenient to decompose the current (\ref{current}) on the eigenoperators of the neutrino Hamiltonian

\begin{equation}
j_{\pm} = 2 U^*_{ei} U^*_{ej} \tau \sigma_{\pm}
\label{operators}
,
\end{equation}
where

\begin{equation}
\sigma_+ =
\left(
\begin{matrix}
0 & 1 \\
0 & 0
\end{matrix}
\right), \ \ \
\sigma_- =
\left(
\begin{matrix}
0 & 0 \\
1 & 0
\end{matrix}
\right)
.
\end{equation}
It should be mentioned that the dissipative operators $j_+$ and $j_-$ in (\ref{operators}) do not commute with the Hamiltonian of the neutrino system, which means that there is the energy dissipation in the neutrino system due to the  emission of a photon.

The equation (\ref{rho}) can be formally solved (integrated). Since we are not interested in the evolution of the electromagnetic field $F$, its degrees of freedom should be traced out

\begin{equation}
\rho_\nu(t_f) = tr_F \left( Texp\left[ \int_{t_i}^{t_f} d^4x  \left[ H(x),\rho(t_i) \right]\right] \right)
\label{w1}
,
\end{equation}
where $\rho_\nu(t) = tr_F \rho(t)$ is the density matrix which describes the evolution of the neutrino system. Note that the trace makes the equation irreversible that leads to appearance of the dissipative terms.

We rewrite equation (\ref{w1}) using the decomposition of the chronological time-ordering operator $T$ into time-ordering operator for a matter current $T^j$ and for electromagnetic fields $T^A$ as $T = T^j T^A$

\begin{multline}
\rho_\nu(t_f) = T^j \left( \exp \left[ \int_{t_i}^{t_f}d^4 x \left[ H_\rho (x),\rho(t_i) \right] \right] \right. \\
\left. tr_F \left\{ T^A \exp \left[ \int_{t_i}^{t_f} d^4x \left[ H_{int} (x),\rho(t_i) \right]  \right]  \right\} \right)
\label{w2}
.
\end{multline}
We exclude time-ordering of the electromagnetic field using Wick theorem \cite{Itzykson}

\begin{widetext}
\begin{multline}
T^A \exp \left[ \int_{t_i}^{t_f} d^4 x \left[ H_{int} (x),\rho(t) \right] \right] = \\
\exp \left[ - \dfrac 1 2 \int_{t_i}^{t_f} d^4 x \int_{t_i}^{t_f} d^4 x' [A_\mu(x), A_\nu(x')][j^\mu(x) j^\nu (x'), \rho (t_i)]  \Theta (t-t') \right] \exp \left[ \int_{t_i}^{t_f} d^4 x [ H_{int} (x),\rho(t_i) ] \right]
\label{w3}
.
\end{multline}
\end{widetext}
 After inserting equation (\ref{w3}) into (\ref{w2}) we get

%%%%%%%%%%%%%%%%%%%%%%%%%%%%%%%%%%%%%%%%%%%%%%%%%%%%%%%%%%%%%%%%%%%%%%%%%%%%%%%%
%\newpage
%%%%%%%%%%%%%%%%%%%%%%%%%%%%%%%%%%%%%%%%%%%%%%%%%%%%%%%%%%%%%%%%%%%%%%%%%%%%%%%%

\begin{widetext}
\begin{multline}
\rho_\nu(t_f) =  T^j  \left(  \exp \left[  \int_{t_i}^{t_f} d^4 x  [H_\rho (x),\rho(t_i)] - \dfrac 1 2 \int_{t_i}^{t_f} d^4 x \int_{t_i}^{t_f} d^4 x' \Theta (t-t') [A_\mu(x), A_\nu(x')] [j^\mu(x) j^\nu (x'), \rho(t_i)] \right]  \right. \times \\
\times \left. tr_f \left\{ \exp \left[ \int_{t_i}^{t_f} d^4 x [ H_{int} (x),\rho(t_i) ] \right] \right\} \right)
\label{w4}
.
\end{multline}
\end{widetext}
Here $\Theta (t) $ is the Heaviside function.
The last term in equation (\ref{w4}) is responsible for the effect of quantum decoherence. In order to calculate this term we will consider the initial state of low entropy that is given by $\rho(t_i) = \rho_\nu(t_i) \otimes \rho_f$ where $\rho_f$ describes the radiation field. Using the cumulant expansion for super operator (see details in \cite{Breuer_Pettrucione}) one can get

\begin{widetext}
\begin{multline}
tr_f \left\{ \exp \left[ \int_{t_i}^{t_f} d^4 x [ H_{int} (x),\rho(t_i) ] \right] \right\} = \\
 \exp \left[ \dfrac 1 2 \int_{t_i}^{t_f} d^4 x \int_{t_i}^{t_f} d^4 x' \{ \langle A_\nu(x') A_\mu(x)\rangle_f j^\mu(x) j^\nu (x') \rho_\nu (t_i) + \langle A_\mu(x) A_\nu(x') \rangle_f \rho_\nu (t_i) j^\mu(x) j^\nu (x') \right. \\
\left. - \langle A_\nu(x') A_\mu(x)\rangle_f j^\mu(x) \rho_\nu (t_i) j^\nu (x') -  \langle A_\mu(x) A_\nu(x') \rangle_f    j^\nu (x') \rho_\nu(t_i) j^\mu(x)\} \right]
\label{w5}
.
\end{multline}
\end{widetext}
The angular brackets denote the average with respect to the radiation field in a thermal equilibrium state at a certain temperature $T_\gamma$:

\begin{equation}
\langle O \rangle_f = tr_f\left(O \frac{1}{Z} \exp [-H_f/k_BT_\gamma]\right)
,
\end{equation}
where $H_f$ represents the Hamiltonian of the free radiation field. In equation(\ref{w5}) we used the fact that $\langle A_\mu \rangle_f = 0$ and that the initial state is Gaussian with regard to the field variables. Inserting (\ref{w5}) into (\ref{w4}) after some algebra gives

\begin{multline}
	\frac{\partial}{\partial t}\rho (t) = -i \left[ H_\nu, \rho(t) \right] - \\
	-\frac{i}{2} \int d^3x \int d^3x' \int_{t_i}^{t_f}  d x_0' D(x-x') \left[\vec{j}(x) , \{\vec{j}(x'), \rho(t) \}  \right] - \\
	-\frac{1}{2} \int d^3x \int d^3x' \int_{t_i}^{t_f}  d x_0' D_1(x-x') \left[\vec{j}(x) , \left[ \vec{j}(x'), \rho(t) \right] \right]
	\label{2-order}
	,
\end{multline}
where

\begin{equation}
D(x-x')_{ij} = i \left[A_i(x), A_j(x')\right]
,
\end{equation}
\begin{equation}
D_1(x-x')_{ij} =  \langle \{A_i(x), A_j(x')\}\rangle_f
\end{equation}
are Pauli-Jordan commutator function and anticommutator function respectively.

Equation(\ref{2-order}) is modified under the assumptions of Marcov and rotating wave approximations. The first approximation consists in the replacement

\begin{equation}
\int_{t_i}^{t}dx_0'\to \int_{-\infty}^{t}dx_0' = \int_{0}^{\infty}d\tau
.
\end{equation}
The rotating wave approximation is equivalent to an averaging procedure over the rapidly oscillating terms.

In the second-order approximation we get the quantum optical master equation for the neutrino case, which is analogous to one for the case of electrons \cite{Breuer_Pettrucione},
\begin{equation}
\dfrac{\partial}{\partial t} \rho_\nu (t) = - i \left[ H_\nu,\rho_\nu(t) \right] - i \left[ H_{S},\rho_\nu(t) \right]+D[\rho_\nu(t)]
.
\label{OpticEquation}
\end{equation}

The Hamiltonian $H_S$ leads to a renormalization of the system Hamiltonian $H_\nu$ which is induced by the vacuum fluctuations of the radiation field and by thermally induced processes. Our goal is to find the dissipative terms, thus we omit the renormalization  part $H_S$ in the following derivations. In (\ref{OpticEquation}) $D(\rho_\nu(t))$ is a dissipator of the equation which can be expressed in the following form (see \cite{Breuer_Pettrucione})

\begin{widetext}
\begin{multline}
D[\rho_\nu(t)] = \frac{\Delta_{ij}}{4 \pi^2}(f(2\Delta_{ij})+1) \left( j_-\rho_\nu (t) j_+ - \frac{1}{2} j_+j_-\rho_\nu(t) - \frac{1}{2} \rho_\nu(t) j_+j_-     \right)+\\
+\frac{\Delta_{ij}}{4 \pi^2} f(2\Delta_{ij}) \left( j_+\rho_\nu (t) j_- - \frac{1}{2} j_-j_+\rho_\nu(t) - \frac{1}{2} \rho_\nu(t) j_-j_+ \right)
,
\label{dissipator}
\end{multline}
\end{widetext}
where $\Delta_{ij}$ is the energy difference between two neutrino mass states $\nu_i$ and $\nu_j$, and $f(E)$ denotes the Planck distribution

\begin{equation}
f(E) = \dfrac{1}{e^{E/kT_\gamma}-1}
,
\end{equation}
where $T_\gamma$ is the temperature of the external photons.

The first term in equation (\ref{dissipator}) is responsible for the spontaneous and the thermally induced emission process and the second one is responsible for the thermally induced absorption process.

In the medium, it is necessary to define new neutrino effective mass states $\tilde{\nu}$ and the effective mixing angle $\tilde{\theta}$. In this new basis the neutrino evolution Hamiltonian is diagonal (see, for example \cite{Pal} and \cite{Freund})  and the energy difference between two neutrino states is expressed as

\begin{equation}
\Delta_{ij} = \dfrac{\sqrt{(\Delta m_{ij} \cos 2\theta_{ij} - A)^2 +\Delta m^2_{ij} \sin^2 2 \theta_{ij}}}{2 E}
,
\end{equation}
where the effective mixing angle is given by

\begin{equation}\label{MSW_effect}
\sin^2 2 \tilde{\theta}_{ij} = \dfrac{\Delta m^2_{ij} \sin^2 2 \theta_{ij}}{(\Delta m_{ij} \cos 2 \theta_{ij} - A)^2 + \Delta m_{ij}^2 \sin^2 2 \theta_{ij}}
,
\end{equation}
where $E$ is the neutrino energy and A is the matter potential. For active-active neutrino oscillations the matter potential is

\begin{equation}
A = 2 \sqrt{2} G_F n_e E
,
\end{equation}
and for the case of active-sterile neutrino oscillations is

\begin{equation}
A = \dfrac{3 \sqrt{2}}{2} G_F n_b (Y_e-\dfrac 1 3)
,
\end{equation}
where $n_b$ is the baryon number density and $Y_e$ is electron fraction.

Putting everything together we get the final expression for neutrino evolution in the effective mass basis

\begin{multline}
\frac{\partial}{\partial t} \rho_{\tilde{\nu}} (t) = - i \left[ H_{\tilde{\nu}},\rho_{\tilde{\nu}}(t) \right] +  \\
+\kappa_1 \left( \sigma_-\rho_{\tilde{\nu}} (t) \sigma_+ - \frac{1}{2} \sigma_+\sigma_-\rho_{\tilde{\nu}}(t) - \frac{1}{2} \rho_{\tilde{\nu}}(t) \sigma_+\sigma_-     \right)+\\
+\kappa_2 \left( \sigma_+\rho_{\tilde{\nu}} (t) \sigma_- - \frac{1}{2} \sigma_-\sigma_+\rho_{\tilde{\nu}}(t) - \frac{1}{2} \rho_{\tilde{\nu}}(t) \sigma_-\sigma_+ \right)
,
\label{Optic2}
\end{multline}
where the Hamiltonian $H_{\tilde{\nu}} = diag(\tilde{E_1},\tilde{E_2})$ and

\begin{equation}
\kappa_1 = \dfrac{\Delta_{ij}}{\pi^2} \sin^2 2\tilde{\theta}_{ij} \tau^2 (f(2\Delta_{ij})+1 )
\label{parameter1}
,
\end{equation}
\begin{equation}
\kappa_2 = \dfrac{\Delta_{ij}}{\pi^2} \sin^2 2\tilde{\theta}_{ij} \tau^2 f(2\Delta_{ij})
\label{parameter2}
\end{equation}
are the parameters that describe decoherence of the neutrino system. For an extreme external environment $f(2\Delta)\gg 1$, thus one can use $\kappa_1 \approx \kappa_2 = \kappa$. Equation (\ref{Optic2}) has the form of Lindblad equation (\ref{LindbladEq}) where decoherence  and relaxation parameters are expressed as

\begin{equation}\label{Gamma_decoherence}
\Gamma_1 = \dfrac \kappa 2
, \ \ \ \
\Gamma_2 = \kappa
.
\end{equation}

The decoherence $\Gamma_1$ and relaxation $\Gamma_2$ parameters   depend on $\sin^2 2 \tilde{\theta} $ that means that the parameters undergo the MSW effect.

It is useful to compare the expression  for the decoherence and relaxation parameters with the rate  of the neutrino radiative decay derived in \cite{Olivio_Nieves_Pal} (see also \cite{Nieves:1997md}). In our case of the transitions between the stationary states we use the results of \cite{Olivio_Nieves_Pal} and \cite{Nieves:1997md} with the substitution $\dfrac{\Delta m_{ij}}{4 E} \to \Delta_{ij}$ and $\sin^2 2 \theta_{ij} \to \sin^2 2 \tilde{\theta}_{ij} $
and get the following expression for the neutrino radiative decay rate in matter
\begin{equation}
\Gamma_r = \dfrac{\Delta_{ij}}{\sqrt{2}} \sin^2 2\tilde{\theta}_{ij} \tau^2 f(2\Delta_{ij})
,
\end{equation}
that coincides with decoherence parameter (\ref{Gamma_decoherence}) up to a constant. This is because the neutrino quantum decoherence appears due to radiative decay.
Therefore, the neutrino quantum decoherence in our consideration can serve
as one of possible indications in favour for the neutrino radiative decay.

The solution of equation (\ref{Optic2}) is given by

\begin{equation}
\rho_{\tilde{\nu}} = \frac 12
\left(
\begin{matrix}
1 + \cos 2 \tilde{\theta}_{ij} e^{-\kappa t} & \sin^2\tilde{2\theta}_{ij}  e^{i 2\Delta_{ij} t} e^{-\kappa t/2} \\
\sin^2\tilde{2\theta}_{ij} e^{-i 2\Delta_{ij} t} e^{-\kappa t/2} &  1 - \cos 2 \tilde{\theta}_{ij} e^{-\kappa t}
\end{matrix}
\right)
\label{solution}
.
\end{equation}
From (\ref{solution}) we obtain the probability of the neutrino flavour oscillations $\nu_e \leftrightarrow \nu_x$ (where $\nu_x$ can be active
$\nu_{\mu, \tau}$ or sterile $\nu_s$ neutrinos)

\begin{multline}
P_{\nu_e \to \nu_x} = \sin^2 2 \tilde{\theta}_{ij} \sin^2\left(\Delta_{ij} x\right) e^{-\kappa x/2} + \\ +
\frac 1 2 \left( 1 - \sin^2 2 \tilde{\theta}_{ij} e^{-\kappa x/2} - \cos^22 \tilde{\theta}_{ij}e^{-\kappa x}\right).
\label{Probability}
\end{multline}	
Here we consider the ultrarelativistic neutrinos and made the substitution $t \to x$.

Nondiagonal elements of the density matrix $\rho_{\tilde{\nu}}$ are responsible for the coherence between the neutrino states $\tilde{\nu}_1$ and $\tilde{\nu}_2$. From (\ref{solution}) it follows that nondiagonal elements are decreasing with the rate $\kappa / 2$ that leads to damping of the amplitude of neutrino oscillations with the same rate. The diagonal elements $\rho_{11}$ and $\rho_{22}$ denote probabilities to find $\tilde{\nu}_1$ and $\tilde{\nu}_2$, respectively. In the limit $t \to \infty$ the neutrino system tends to the thermal equilibrium which gives the oscillation probability $P_{\nu_e \to \nu_x} \to \frac 1 2$.\\

Obviously, the possibility for the direct detection of the neutrino radiative decay seems to be shadowed due to high luminosity of the astrophysical external environment (in the terrestrial conditions the decay is highly suppressed). However, we have shown that the radiative decay can also influence the neutrino oscillations probability and therefore can modify the neutrino spectrum. This influence can be detected or constrained in the present and future neutrino experiments that will give information about the neutrino radiative decay. Since there are still uncertainties in the neutrino characteristics (e.g., in the value of the neutrino magnetic moment) and there are theories that predict neutrino non-standard interaction the information on the neutrino radiative decay can provide us new constrains on physics beyond the Standard Model. As an example, in the next section we consider the  influence of the massless dark photons on the neutrino quantum decoherence.

It should be noted that formula (\ref{Probability}) is valid for the neutrino evolution in a constant or an adiabatically changing density of the environment. The case of nonadiabatic neutrino oscillations needs to be treated separately. In addition, in certain  astrophysical  environments (e.g., supernovae bursts) one should also take into account the collective neutrino oscillations.

%%%%%%%%%%%%%%%%%%%%%%%%%%%%%%%%%%%%%%%%%%%%%%%%%%%%%%%%%%%%%%%%%%%%%%%%%%%%%%%%
\section{neutrino quantum decoherence engendered by neutrino radiative decay to a dark photon}
%%%%%%%%%%%%%%%%%%%%%%%%%%%%%%%%%%%%%%%%%%%%%%%%%%%%%%%%%%%%%%%%%%%%%%%%%%%%%%%%%

The quantum decoherence can also be a result of  possible interaction between neutrino and dark matter. Here below we generalize the developed above approach to describe the effect of the quantum decoherence engendered by the neutrino radiative decay to a massless dark photon. The  massless dark photon is a boson field of a new gauge symmetry $U(1)_X$ \cite{Babu:1997st,Foot:1991kb,Gang_Li}. This boson does not interact directly with ordinary matter, but there is a mixing between dark photons and photons of the standard model. This can induce interactions between dark photons and the SM particles. Since the new gauge boson field $X$ is similar to the gauge boson field in the standard model, the vertex of the neutrino radiative decay can be written in the same way as it is in  (\ref{vertex})

\begin{equation}
\Gamma^X_\alpha = U^*_{ei} U_{ej} \tau^X_{\alpha\beta} \gamma^\alpha L
,
\end{equation}
where for the massless dark photons

\begin{equation}
\tau_{\alpha\beta}^{X} =\tau_{X} P_{\alpha\beta} = - \frac{\sigma c_W}{\sqrt{1-\sigma^2c_W^2}} \frac{e G_F T^2}{2 \sqrt{2}} P_{\alpha\beta}
\label{Xtau}
,
\end{equation}
 and $c_W = \cos \Theta_W$ is the cosine of the weak mixing angle, $\sigma$ characterises the mixing between dark photons and photons of the standard model. Here we consider only the case of the
extreme relativistic electrons. Then the neutrino decoherence parameter in case the neutrino radiative decay to a dark photon can be written as

\begin{equation}\label{dark_photon}
\kappa_X = \dfrac{\Delta_{ij}}{\pi^2} \sin^2 2\tilde{\theta}_{ij} \tau_X f_X(2\Delta_{ij})
,
\end{equation}
where

\begin{equation}
f_X(E) = \dfrac{1}{e^{E/kT_X}-1}
\end{equation}
is the Planck distribution for dark photons which temperature is $T_X$. Thus, the neutrino oscillation probability (\ref{Probability}) can be modified by the neutrino interactions with dark photons via the effect of the quantum decoherence characterized by (\ref{dark_photon}).

%%%%%%%%%%%%%%%%%%%%%%%%%%%%%%%%%%%%%%%%%%%%%%%%%%%%%%%%%%%%%%%%%%%%%%%%%%%%%%%%%
\section{Quantum decoherence in astrophysical media }
%%%%%%%%%%%%%%%%%%%%%%%%%%%%%%%%%%%%%%%%%%%%%%%%%%%%%%%%%%%%%%%%%%%%%%%%%%%%%%%%%

Most readily the effect of the quantum decoherence in neutrino oscillations can manifest itself in extreme astrophysical environments with extremely  relativistic electrons characterized by high temperatures.
In this case the neutrino electromagnetic interactions with photons is given by (\ref{vertex}) and   (\ref{ER}). The decoherence and relaxation parameters and the corresponding effect of quantum decoherence depend significantly upon the electron temperature $T$.  Sufficiently high temperatures arise during supernovae bursts, where the electron and photon temperatures can reach values up to 30 MeV and 100 MeV, respectively \cite{Bolling_Janka_Lohs}. Therefore, we consider the proposed mechanism of quantum decoherence in supernovae (SN) environments.

\begin{figure}[h]
	\centering
	\includegraphics[width=0.8 \linewidth]{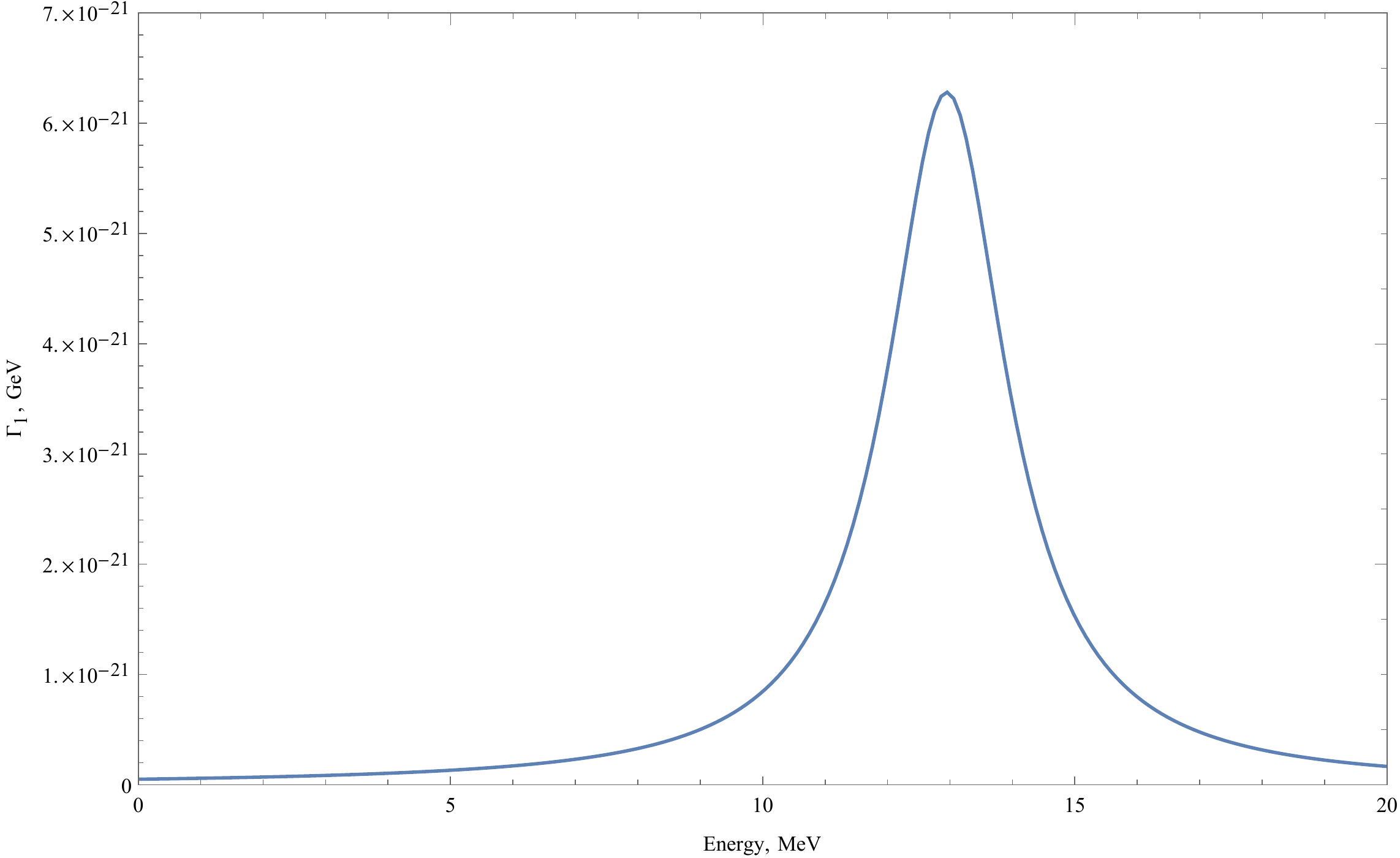}
	\caption{\label{Parameter}}{The energy dependence of the neutrino decoherence parameter $\Gamma _1$ (\ref{Gamma_decoherence}) in case of active-sterile neutrino oscillations $\nu_e \leftrightarrow \nu_s$  in a supernova invironment.}
\end{figure}

From equations (\ref{parameter1}), (\ref{parameter2}) and (\ref{MSW_effect})  one can see that the effect of the quantum decoherence undergoes the resonance similar to the MSW effect. Therefore, the effect of neutrino quantum decoherence is maximal in the resonance regions where $\sin^2(2\tilde{\theta}_{ij}) = 1$. For the active-sterile neutrino oscillations $\nu_e \leftrightarrow \nu_s$ the MSW effect occurs much closer  to the center of the SN explosion \cite{Tamborra_Janka,Wu:2013gxa} where highest  temperatures are expected. For this case the maximal value of the decoherence parameter is of order $\Gamma_1 \sim 10^{-21}$ GeV at the resonance region  (see Fig.\ref{Parameter}). In case of active-active neutrino oscillations $\nu_e \leftrightarrow \nu_{\tau}$ the parameter is significantly smaller (of order of $\Gamma_1 \sim 10^{-31} $) GeV.

\section{Conclusion}
A new theoretical framework, based on the quantum field theory of open systems applied to neutrinos, has been developed to describe the neutrino evolution in external environments accounting for the effect of the neutrino quantum decoherence.

We have used this approach to consider a new mechanism of the neutrino quantum decoherence engendered by the neutrino radiative decay to photons and dark photons in the thermal background of electrons. We have obtained the explicit expressions of the decoherence and relaxation parameters as functions of the characteristics of an external environment and also of the neutrino energy and calculated the corresponding neutrino oscillation probabilities $P_{\nu_e \to \nu_x}$. Note that the influence of the neutrino radiative decay, as well as the
neutrino interaction with dark photons,  on the neutrino oscillation phenomenon
and the corresponding contributions to the neutrino quantum decoherence are
considered for the first time.

An estimation of the effect of the neutrino quantum decoherence has been given for a particular case of a supernova explosion. The obtained values of the decoherence parameter for neutrino active-sterile oscillations $\nu_e \leftrightarrow \nu_s$ are of order of $\Gamma_1 \sim 10^{-21}$ GeV and $\Gamma_1 \sim  10^{-31}$ GeV for neutrino active-active oscillations $\nu_e \leftrightarrow \nu_{\tau}$. To estimate the scale of the obtained values, we compare our results, derived for the dense astrophysical environment, with those for different environments known in literature: $\Gamma_1 \sim 10^{-24}$ GeV from the reactor neutrino fluxes \cite{Coloma} and $\Gamma_1 \sim 10^{-28}$ GeV from the solar neutrino fluxes \cite{deHolanda:2019tuf}.

The developed theoretical framework provides a  general basis for the detailed description of the neutrino quantum decoherence due to different neutrino interactions with external environments. The importance of this approach is highlighted by the prospects of the forthcoming new large volume neutrino detectors (such as, for instance, JUNO, DUNE and Hyper-Kamiokande) that will provide new frontier in high-statistics measurements of neutrino fluxes from supernovae.

The studies and presented results of this paper are also important due to the fact that nonstandard neutrino interactions are also one of the possible sources of the neutrino quantum decoherence. Therefore, implementation of the developed theoretical framework for description of the neutrino quantum decoherence to analysis of the astrophysical and terrestrial neutrino fluxes can open a new window to check physics beyond the standard model.

\section{Acknowledgements}

The authors are thankful to Alexander Grigoriev for his helpful comments on the manuscript and to Konstantin Kouzakov for valuable discussions. This work was supported by the Russian Foundation for Basic Research under grants No. 20-52-53022-GFEN-a.

%\begin{figure}[h!]
%	\centering
%	\includegraphics[width=0.8 \linewidth]{Oscillations.pdf}
%	\caption{\label{Oscillations}}{ The probability of the active to sterile neutrino oscillations $\nu_e \to \nu_s$ in supernovae taking into account the quantum decoherence (the neutrino energy is $E = 15\ \ MeV$).}
%\end{figure}

%\begin{figure}[h!]
%	\begin{minipage}[h]{0.49\linewidth}
%		\center{\includegraphics[width=1\linewidth]{SpectrumSTD} \\ a)}
%	\end{minipage}
%	\hfill
%	\begin{minipage}[h]{0.49\linewidth}
%		\center{\includegraphics[width=1\linewidth]{SpectrumDec} \\ b)}
%	\end{minipage}
%	\caption{\label{Spectrum}}{The energy dependence of the $\nu_e \to \nu_s$ transition probabilities at the distance of 50 km from the neutrinosphere , a) is for standard oscillations with no quantum decoherence, and b) is for oscillations taking into account the quantum decoherence.}
%\end{figure}

\end{document}